\documentstyle[11pt,epsfig]{article}
\begin{document}

\thispagestyle{empty}
\begin{flushright}
{CERN-TH/98-321}\\
{FTUAM-98-19}
\end{flushright}
\vspace*{1cm}
\begin{center}
{\large{\bf Neutrino oscillation physics with a neutrino factory} }\\
\vspace{.5cm}
A. De
  R\'ujula$^{\rm a,}$\footnote{derujula@nxth21.cern.ch},
M.B. Gavela$^{\rm b,}$\footnote{gavela@garuda.ft.uam.es,}
and 
P. Hern\'andez$^{\rm a,}$\footnote{pilar.hernandez@cern.ch. On leave 
from Dept. de F\'{\i}sica Te\'orica, Universidad de Valencia.}

\vspace*{1cm}
$^{\rm a}$ Theory Division, CERN,
  1211 Geneva 23, Switzerland\\
$^{\rm b}$ Dept. de F\'{\i}sica Te\'orica, Univ. Aut\'onoma de
Madrid, Spain

\end{center}
\vspace{.3cm}
\begin{abstract}
\noindent
Data from atmospheric and solar neutrinos indicate
that there are at least three neutrino types involved
in oscillation phenomena. Even if the corresponding
neutrino mass scales are very different, the inevitable
reference to mixing between more than two neutrino
types has profound consequences on the planning of
the accelerator experiments suggested  by these results.
We discuss the 
measurement of mixing angles and CP phases in the
context of the neutrino beam emanating
from a {\it neutrino factory}: the straight sections
of a muon storage ring. We emphasize the importance
of charge identification. The appearance of wrong sign
muons in a long baseline experiment may 
provide a powerful test of neutrino oscillations in the
mass-difference range indicated by atmospheric-neutrino
observations.
\end{abstract}

\vspace{5cm}
\begin{flushleft}
{CERN-TH/98-321}\\
{FTUAM-98-19}\\
\end{flushleft}
\newpage

\section{Motivation in the current situation}
\bigskip
Recently published strong indications of
atmospheric neutrino oscillations
\cite{Superka}
have rekindled the interest in accelerator experiments
that could study the same range of 
parameter space. The results of SuperKamiokande
are interpreted as oscillations of muon neutrinos into
neutrinos that are not $\nu_e$s. Roughly
speaking, the measured mixing
angle is close to maximal: $\sin^2 2 \theta > 0.8$,
and $\Delta m^2$ is in the range $5\times 10^{-4}$ to $6\times 10^{-3}$
eV$^2$, all at 90\% confidence. 

The SuperK mass (squared) range is one order
 of magnitude below the previous Kamiokande observations
\cite{ka}, just what is needed to render the oft-discussed
long baseline experiments --such as MINOS \cite{MINOS}
or a CERN to Gran Sasso \cite{CGS}
project-- hardly capable of covering the whole parameter range
of interest.

The solar neutrino deficit is interpreted either as MSW (matter
enhanced) oscillations \cite{MSW} or as vacuum oscillations 
\cite{osc} that deplete
the original $\bar \nu_e$s, presumably in favour of 
$\bar \nu_\mu$s. The corresponding mass differences 
--$10^{-5}$ to $10^{-4}$ eV$^2$ or some $ 10^{-10}$ 
eV$^2$--
are significantly below the range deduced from
atmospheric observations. Currently discussed terrestrial
experiments have no direct access to the solar mass range(s).

A straight section in a high intensity muon storage ring is
an excellent putative source of neutrinos \cite{muring}:
a {\it neutrino factory}. 
The normalization, energy and angular spectrum of the
$\nu_\mu+\bar\nu_e$ or $\bar\nu_\mu+\nu_e$
beams would be known to
very high precision. The relative amounts of (forward-moving)
electron neutrinos can be tuned by varying the muon
polarisation.
With a very intense but not
unrealistic proton accelerator (with some 100 times the 
current of the current CERN-PS)
it is possible to dream of neutrino beams two orders of 
magnitude more intense than existing ones \cite{muring,Dydak}.

For the sake of illustration, we shall consider as
a {\it ``reference set-up''} the neutrino
beams resulting from the decay of $n_\mu=2\times 10^{20}$ $\mu^+$s
and/or $\mu^-$s in a straight section of an $E_\mu=20$ GeV
muon accumulator ring pointing at an experiment with
a 10 kT target, some 732 km downstream,
roughly the distance from CERN to Gran Sasso or from
Fermilab to the Soudan Lab. Most of our figures are for
the  ``reference baseline'' $L=732$ km, but
we specify the scaling
laws that relate the results at different energies and distances.
When considering
the possibility of detecting the production of $\tau$s
we use the example of the Opera proposal \cite{Opera}:
a one kTon target with a $\tau$-detection efficiency 
(weighed with the branching ratio of observable channels) of 35\%.

Appearance experiments (e.g. $\tau$ production in a 
high-energy beam from
$\mu$ decay) are more sensitive and potentially more convincing
than disappearance experiments.
Given the current solar and atmospheric results, one must
unavoidably analyze the prospects of neutrino oscillations
in a neutrino factory in a three-generation mixing scenario.
As it turns out, this scenario brings to the fore the importance
of appearance channels other than $\tau$ production, e.g.,
the production of ``wrong sign'' muons, a channel for which
there would be no beam-induced background at a neutrino factory.
We discuss the physics backgrounds in Chapter 6, rather
briefly, as we cannot embark on a more thorough discussion of this 
issue without a specific detector in mind.

Our emphasis is not on the traditional and very well studied
 $\tau$-appearance channel, but on the wrong sign muons,
which are more specific to a neutrino factory. We choose the
most conservative scenario regarding the neutrino masses:
$\Delta m_{23}^2$ is given by the SuperK observations, 
and $\Delta m_{12}^2$ by the ensemble of solar experiments
(disregarding one of the latter or accepting the results of LSND
\cite{LSND}
opens the way to larger mass differences and oscillatory signals). 

We devote the next Section to a two-by-two mixing scenario
in order to illustrate the differences with the three-by-three
case, to which we return thereafter.

\section{Generalities in a two-family context}
\bigskip

Interpret the atmospheric neutrino data as 
$\nu_\mu \leftrightarrow \nu_\tau$ oscillations with
a mixing angle $\sin^2 ( 2 \theta_{23}) \sim 1$ and
$5\times 10^{-4}$ 
eV$^2$ 
$<\Delta m_{23}^2<6\times 10^{-3}$
eV$^2$. 
In a two-family scenario the oscillation probability is:
\begin{equation}
P(\nu_\mu\rightarrow\nu_\tau)=
\sin^2 (2 \theta_{23})\,
\sin^2\left({\kappa\,\Delta m_{23}^2\,L\over E_\nu}\right) \, ,
\label{twofamprob}
\end{equation}
with $\kappa=1/4$ in natural units or $\kappa=1.27$ in
GeV per km and eV$^2$. 

The mass splitting regions $\Delta m_{12}^2$
preferred by solar neutrino observations are such that
$\Delta m_{12}^2\,L/E_\nu$ would be very small in long baseline
experiments on Earth. If $\nu_e \leftrightarrow \nu_\mu$
oscillations are described by the $(23)\to (12)$ analogue of
Eq.(\ref{twofamprob}), oscillations between the first two
generations would be unobservable in terrestrial experiments.
Though well known to be an oversimplification \cite{FL},
a mixing of two generations at a time is often
assumed, leading to potentially misleading conclusions.   

With
stored $\mu^-$s one has a $\nu_\mu+\bar \nu_e$ beam.
The observable $\nu_\mu\to \nu_\tau$ oscillation signals are:
\begin{eqnarray}
 \mu^- \rightarrow e^-\,  & \nu_\mu &  \, \bar{\nu}_e\, ;
 \nonumber\\
& \;  &\bar{\nu}_e  \rightarrow \bar{\nu}_e \rightarrow e^+ \;\; {\rm normalization,}
\nonumber\\
&  \nu_\mu  &  
\rightarrow \nu_\mu\rightarrow \mu^- \;\;\;\;\; {\rm disappearance,}
\nonumber\\
& \nu_\mu & \rightarrow \nu_\tau \rightarrow \tau^- \;\;\;\;\; {\rm appearance.}
\label{nocharges}
\end{eqnarray}
In the absence of dominant backgrounds, the statistical sensitivity
--that we define throughout as the smallest effect
that can be excluded with 90\% confidence--
is very different for appearance and disappearance
processes. In the case of $\nu_\mu$-disappearance
and for $N_\mu$  expected
events, the fractional sensitivity in the measurement
 of
 the flux- and cross-section weighed probability  
$\bar P(\nu_\mu\rightarrow\nu_\tau)$ is 
$1.65/\sqrt{N_\mu}$.
For $\nu_\tau$ appearance, there being no $\nu_\tau$ 
contamination in the beam, the non-observation of 
$\tau$ events would establish a 90\% limit
$\bar P(\nu_\mu\rightarrow\nu_\tau)<2.44/N_\tau$, with
$N_\tau$ the number of events to be expected, should all
$\nu_\mu$s be transmogrified into $\nu_\tau$s.

The neutrino fluxes at a neutrino factory have simple analytical 
forms\footnote{We expect   the
$\nu$ beam divergence to be dominated by the $\mu$-decay kinematics \cite{muring}.}.
Let $y=E_\nu/E_\mu$ be the fractional neutrino energy.
For unpolarized
muons of either polarity, and neglecting corrections of order
$m_\mu^2/E_\mu^2$, the normalized fluxes of forward-moving
neutrinos are:
\begin{eqnarray}
F_{\nu_\mu,\bar\nu_\mu}(y) &\simeq& 2 \,  y^2 \, (3-2 y)
\Theta(y)\,\Theta(1-y)\, ,
\cr
F_{\nu_e,\bar\nu_e}(y) &\simeq& 12\,y^2\, (1- y) 
\Theta(y)\,\Theta(1-y) \, ,
\end{eqnarray}
and, for each produced neutrino type, the forward flux 
from $n_\mu$ $\mu$-decays is:
\begin{equation}
{dN_\nu\over dy\, dS}\Bigm|_{\theta \simeq 0}
\simeq{E^2_\mu\; n_\mu\over \pi\, m_\mu^2\,L^2}\;F_\nu (y)\; .
\label{flux}
\end{equation}
The above expressions are valid at a forward-placed detector of 
transverse dimensions much smaller than the beam aperture.

In the absence of oscillations
one can use Eq.(\ref{flux}) and the charged-current
inclusive cross sections per nucleon on an approximately
isoscalar target ($\sigma_\nu\sim 0.67\times 10^{-38}\, E_\nu$
cm$^2$/GeV, $\sigma_{\bar\nu}\sim 0.34\times 10^{-38}\, E_{\bar\nu}$
cm$^2$/GeV \cite{Boehm})
to compute the number of neutrino interactions. 
For the reference set-up and baseline defined in Section 1,
one expects some $2.2\times 10^{5}$ 
$\mu^-$ ($1.1\times 10^{5}$ 
$\mu^+$) and $9.6\times 10^4$ $e^+$ ($1.9\times 10^5$ $e^-$)
events in a beam
from $\mu^-$ ($\mu^+$) decay \cite{muring}. In our calculations
we make a cut $E_\nu>5$ GeV to eliminate inefficiently observed
low energy interactions. This affects the quoted numbers only
at the few per cent level.

\begin{figure}[htb]
  \centering
\mbox{\epsfig{file=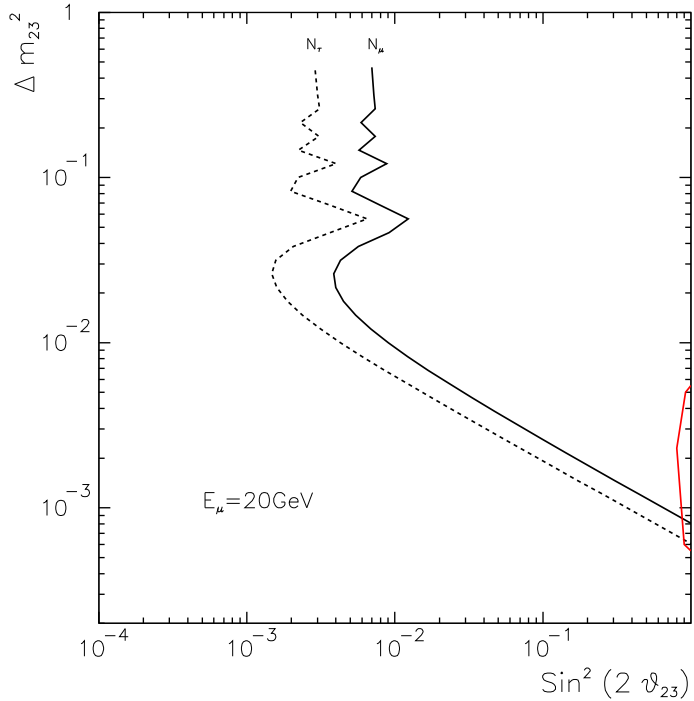,width=4.in,height=4.in}}
  \caption{Sensitivity reach in the [$\sin^2 (2 \theta_{23}),\Delta m_{23}^2$]
plane, at 90\% confidence, for our reference beam and detectors
and $L=732$ km.
Continuous (dashed) boundaries correspond to $\mu$ disappearance
($\tau$ appearance). The small region close to $\sin^2 (2 \theta_{23})=1$
is the SuperK domain.} 
\label{fig:mutau}
\end{figure}

In Fig.~\ref{fig:mutau} we show the   
[${\sin^2(2 \theta_{23}),\Delta m_{23}^2}$] sensitivity to $\mu$-disappearance,
basing it on the measurement of $N_\mu$, the total (energy-integrated) 
number of muon
events. 
By assumption,
$\nu_e$s do not observably oscillate over terrestrial baselines
so that,
in a two-generation scenario, the results would be identical
if extracted from the ratio $N_\mu/N_e$, as in a recent
discussion of an experiment at a $\nu$-factory \cite{bcr1}.
For a $\tau$ detector we refer to Opera \cite{Opera},
in its version described in the introduction. With use
of the cross section $\sigma(\nu_\tau\to \tau)$ given in
\cite{AlbJar}, we also report the 
$\tau$-appearance statistical sensitivity in Fig.~\ref{fig:mutau}, basing it
on the expectation for $N_\tau/N_\mu$. In practice, the search for
$\tau$ events is affected by a steadfast charm background.

For our reference beam, detector and baseline,
the moral from this brief analysis of the two-family scenario is that a 10 kTon experiment capable of telling muons
from electrons (or from neutral currents) would be insufficient
to cover the SuperK mass range. The smaller detector we considered, capable
of telling $\tau$ events from the rest, would also barely suffice.
 
To study the oscillatory signal
of the two-family scenario of Eq.(\ref{nocharges}) there
is no advantage in
measuring the charges of the produced charged leptons: 
for a stored
$\mu^-$ beam one expects charged current neutrino
interactions leading to positrons
and negatively charged heavier leptons, as in Eq.(\ref{nocharges}).            For Majorana neutrinos this is not strictly correct, but 
the specific wrong-sign and CP-violating effects are 
suppressed by an unsurmontable factor $m_\nu/E_\nu$. 
In a three-neutrino mixing scenario, contrarywise,
measuring charges could
be extremely useful and CP-violation effects are not suppressed
by the mentioned factor.

\section{Three-family mixing.}
\bigskip

The mixing between $\nu_e$, $\nu_\mu$ and $\nu_\tau$
is described by a conventional Kobayashi-Maskawa matrix
$V$ relating flavour to mass eigenstates
(we are assuming throughout this note 
that {\it neutrino fluctuat nec mergitur:} there are no 
transitions to sterile neutrinos).
For Dirac neutrinos\footnote{For Majorana neutrinos 
fewer phases are reabsorbable by 
field redefinitions and the mixing matrix is of the form
$V'=V\;V_{_{\rm{M}}}$ with 
$V_{_{\rm{M}}}=\rm{Diag}\, (e^{i\alpha},e^{i\beta},1)$. The effects
of these extra phases are of order $m_\nu/E_\nu$.}
and in an obvious notation:
\begin{equation}
\left(\matrix{\nu_e \cr \nu_\mu \cr\nu_\tau}\right)
= \left(\matrix{
c_{12}c_{13} & c_{13}s_{12} & s_{13} \cr
-c_{23}s_{12}e^{i\delta} -c_{12}s_{13}s_{23}
& c_{12}c_{23}e^{i\delta} -s_{12}s_{13}s_{23}
 & c_{13}s_{23} \cr
s_{23}s_{12}e^{i\delta} -c_{12}c_{23}s_{13}
& -c_{12}s_{23}e^{i\delta} -c_{23}s_{12}s_{13}
 & c_{13}c_{23} \cr}
\right)
\left(\matrix{\nu_1 \cr \nu_2 \cr\nu_3}\right).
\label {CKM}
\end{equation}
Without loss of generality, we choose the convention in
which all Euler angles lie in the first quadrant:
$0<\theta_{ij}<\pi/2$, while the CP-phase is unrestricted:
$0<\delta<2\,\pi$.
Define 
\begin{equation}
W_{\alpha\beta}^{jk}\equiv  \,
[V_{\alpha j}V_{\beta j}^* V_{\alpha k}^*V_{\beta k}]
\end{equation}
and
\begin{equation}
\Delta_{jk}\equiv\frac{ \Delta m_{jk}^2}{2\,E_\nu}\; .
\label{Deltas}
\end{equation}
The transition probabilities between
different flavours are:
\begin{equation}
P(\nu_\alpha\rightarrow \nu_\beta)\,=\, 
-4\; \sum_{k>j}\,{\rm Re}[W_{\alpha\beta}^{jk}]\, 
\sin^2\left({\Delta_{jk}\,L\over 2}\right)
\,\pm\, 2  \,
\sum_{k>j}\, {\rm Im}[W_{\alpha\beta}^{jk}]\, \sin(\Delta_{jk}\,L)
\label{reim}
\end{equation}
with the plus (minus) sign referring to neutrinos (antineutrinos).

Let us adopt, from solar and atmospheric experiments,
the indication that
$|\Delta m_{12}^2|\ll |\Delta m_{23}^2|$, that Barbieri
{\it et al.} \cite{Barb} have
dubbed the ``minimal scheme''. Though this mass hierarchy may
not be convincingly established, the minimal scheme suffices 
for our purpose of delineating the main capabilities of
a $\nu$ factory (we have to deviate from minimality only
in the discussion of CP violation). 

The difference between neutrino
propagation in vacuum and in matter turns out not to have an
important effect on the sensitivity limits that we discuss in this
chapter (for a fixed baseline of $732$ km). They are relevant at larger
distances. We postpone their discussion to the next chapter,
though the figures introduced anon do take the
matter effects into account.

Atmospheric or terrestrial experiments have
an energy range such that $\Delta m^2\, L/E_\nu\ll 1$ for the smaller
($\Delta m_{12}^2$)
but not necessarily for the larger ($\Delta m_{23}^2$) of these mass gaps.
Even then, solar and atmospheric (or terrestrial) experiments 
are not (provided 
$\theta_{13}\neq 0$)
two separate two-generation mixing effects. In the minimal scheme 
solar effects
are accurately described by three parameters 
($\theta_{12}$, $\Delta m_{12}^2$ and $\theta_{13}$),
while the terrestrial effects of interest here depend on
$\theta_{23}$, $\Delta m_{23}^2$ and $\theta_{13}$:
\begin{eqnarray}
P(\nu_e\rightarrow\nu_\mu)&=&  \sin^2(\theta_{23})\, 
\sin^2(2\theta_{13})\,\sin^2\left({\Delta_{23}\, L\over 2}\right) 
\cr
P(\nu_e\rightarrow\nu_\tau)&=&   \cos^2(\theta_{23})\, 
\sin^2(2\theta_{13})\,\sin^2\left({\Delta_{23}\, L\over 2}\right)
\cr
P(\nu_\mu\rightarrow\nu_\tau)&=&  \cos^4(\theta_{13})\, 
\sin^2(2\theta_{23})\,\sin^2\left({\Delta_{23}\, L\over 2}\right)\; .
\label{todasprobs}
\end{eqnarray}
In the minimal scheme CP and T violation effects can be neglected,
so that $P(\bar\nu_\alpha\to\bar\nu_\beta)=P(\nu_\alpha\to\nu_\beta)$
and $P(\nu_\beta\to\nu_\alpha)=P(\nu_\alpha\to\nu_\beta)$. With this
information, Eqs.(\ref{todasprobs}) and unitarity one can construct
all relevant oscillation amplitudes, e.g. $P(\nu_\mu\to\nu_\mu)$.

The approximate analysis of the SuperK data 
by Barbieri {\it et al.}
\cite{Barb} results (for the range of
$\Delta m_{23}^2$ advocated by the SuperK collaboration) 
in the restrictions $\theta_{23}=45\pm 15^o$ and
$\theta_{13}\sim 0\div 45^o$, with a preferred value
around 13$^o$.  Fogli {\it et al.} conclude \cite{Fogli},
after a more thorough analysis and with equal
confidence, that $\theta_{13}<23^0$, while their range
of $\Delta m_{23}^2$ is a little narrower than the one 
obtained by the SuperK team \cite{Superka}.
We shall present results for the range of angles advocated in \cite{Barb}
and the range of masses of \cite{Superka},
simply because they are the widest.

All mixing probabilities in Eq.(\ref{todasprobs}) 
have the same sinusoidal dependence
on $\Delta m^2_{23}\, L/E_\nu$, entering into the description of a plethora of 
channels:
\begin{eqnarray}
 \mu^- \rightarrow e^-\,  & \nu_\mu &  \, \bar{\nu}_e\, ;
 \nonumber\\
& \; & \bar{\nu}_e  \rightarrow \bar{\nu}_e \rightarrow e^+ \;\; {\rm disappearance,}
\nonumber\\
& \; & \bar{\nu}_e  \rightarrow \bar{\nu}_\mu \rightarrow \mu^+ \;\; {\rm appearance,}
\nonumber\\
& \; & \bar{\nu}_e  \rightarrow \bar{\nu}_\tau \rightarrow \tau^+ \;\; {\rm appearance}
\;\;\; (\tau^+ \rightarrow \mu^+;\; e^+)\, ,
\nonumber\\
&  \nu_\mu  &  
\rightarrow \nu_\mu\rightarrow \mu^- \;\;\;\;\; {\rm disappearance,}
\nonumber\\
& \nu_\mu & \rightarrow \nu_e \rightarrow e^- \;\;\;\;\; {\rm appearance,}
\nonumber\\
& \nu_\mu & \rightarrow \nu_\tau \rightarrow \tau^- \;\;\;\;\; {\rm appearance}
\;\;\; (\tau^- \rightarrow \mu^-;\, e^-)\, .
\label{charges}
\end{eqnarray}
The wrong sign channels of $\mu^+$, $\tau^+$ and $e^-$ appearance
are the good news, relative to the two-generation analysis
of Eqs.(\ref{nocharges}).

We extract results on the sensitivity to oscillations
from observable numbers 
of muons, and not from ratios such as the number of muons 
upon the number of electrons, that are so useful in the analysis of
atmospheric neutrinos. Our conclusions would be essentially
identical, were we to draw them from the customary ratios.
Yet, we refer directly to muon numbers not only because the
neutrino-factory flux would be very well understood
(obviating the main reason to take ratios), but also
because the physics of three-generation mixing leads us to advocate
the advantages of an experiment capable of measuring the
charge of muons. It is likely that a relatively large experiment
of this kind would compromise the possibility of efficiently 
distinguishing electron- from neutral-current events. 
Naturally, a complementary experiment on the same beam,
capable of observing electrons with precision, would be useful
\cite{CR,Thom}.

\begin{figure}[htb]
\centering
\mbox{\epsfig{file=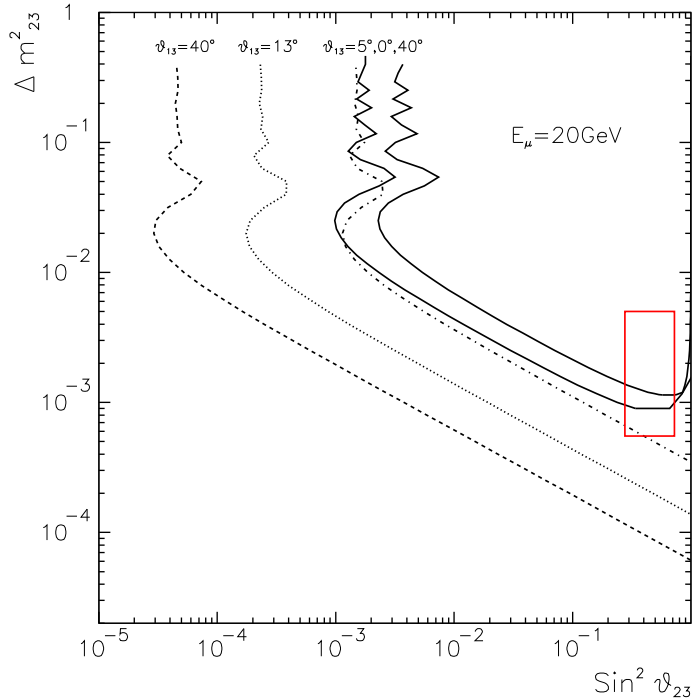,width=4.in,height=4.in}}
  \caption{Sensitivity reach in the plane 
$[\sin^2 \theta_{23},\Delta m_{23}^2]$
 at 90\% confidence, for our reference set-up, a $\mu^-$-decay
beam and $L=732$ km. Matter effects are taken into account. 
The discontinuous lines
correspond to the appearance observable
$N[\mu^+]$ (at $\theta_{13}=40,13,5^0$) and
the full lines correspond to the disappearance
observable $N_\mu$ at $\theta_{13}=0,40^0$.
The rectangle is the approximate domain allowed by 
SuperK data.} 
\label{fig:t23m}
\end{figure}

In Fig.~\ref{fig:t23m} we show the sensitivity reach, in the 
[$\sin^2 (\theta_{23}),\Delta m_{23}^2$] plane
for various values of $\theta_{13}$, for $L=732$ km,
for our reference set-up and for stored $\mu^-$s. 
We have chosen to
illustrate the disappearance observable 
$N_\mu\equiv N[\mu^+ +\mu^-]$ and the appearance measurement $N[\mu^+]$ (the effects of the small $\mu^+$ contamination from
$\bar\nu_e\to\bar\nu_\tau$ oscillations, $\tau^+$ production and 
$\tau^+\to\mu^+$ decay, are negligible). 
Figure~\ref{fig:t23m} conveys an important point: for stored
$\mu^-$s the observation of $\mu^+$  appearance
is very superior to a measurement (such as the depletion
of the total number of muons) in which
the charges of the produced leptons are not measured.
This is true for all $\theta_{13}$  bigger than a few degrees.
This angle is very unconstrained by current measurements. 
Notice that the SuperK domain would be covered for any
$\sin^2(\theta_{13})> 3.6 \times 10^{-3}$ by the appearance channel, while the disappearance
measurement would fall short of this motivating goal.
All these statements refer to statistical sensitivities, in the
absence of the backgrounds discussed in Section 6.
 
Fig.~\ref{fig:mutau} and its comparison with
Fig.~\ref{fig:t23m} 
convey our point regarding the benefits of muon-charge identification.
We are showing results only for stored $\mu^-$s. The 
wrong-sign muon results are slightly superior for the polarity
we do not show: if it is positive, and for equal numbers of decays,
the unoscillated numbers of expected 
electron events (and of potential wrong-sign muons) are
roughly twice as numerous. The $\mu$-disappearance results, on
the other hand, are slightly weaker for a $\mu^+$ beam.

\begin{figure}[htb]
\centering
\mbox{\epsfig{file=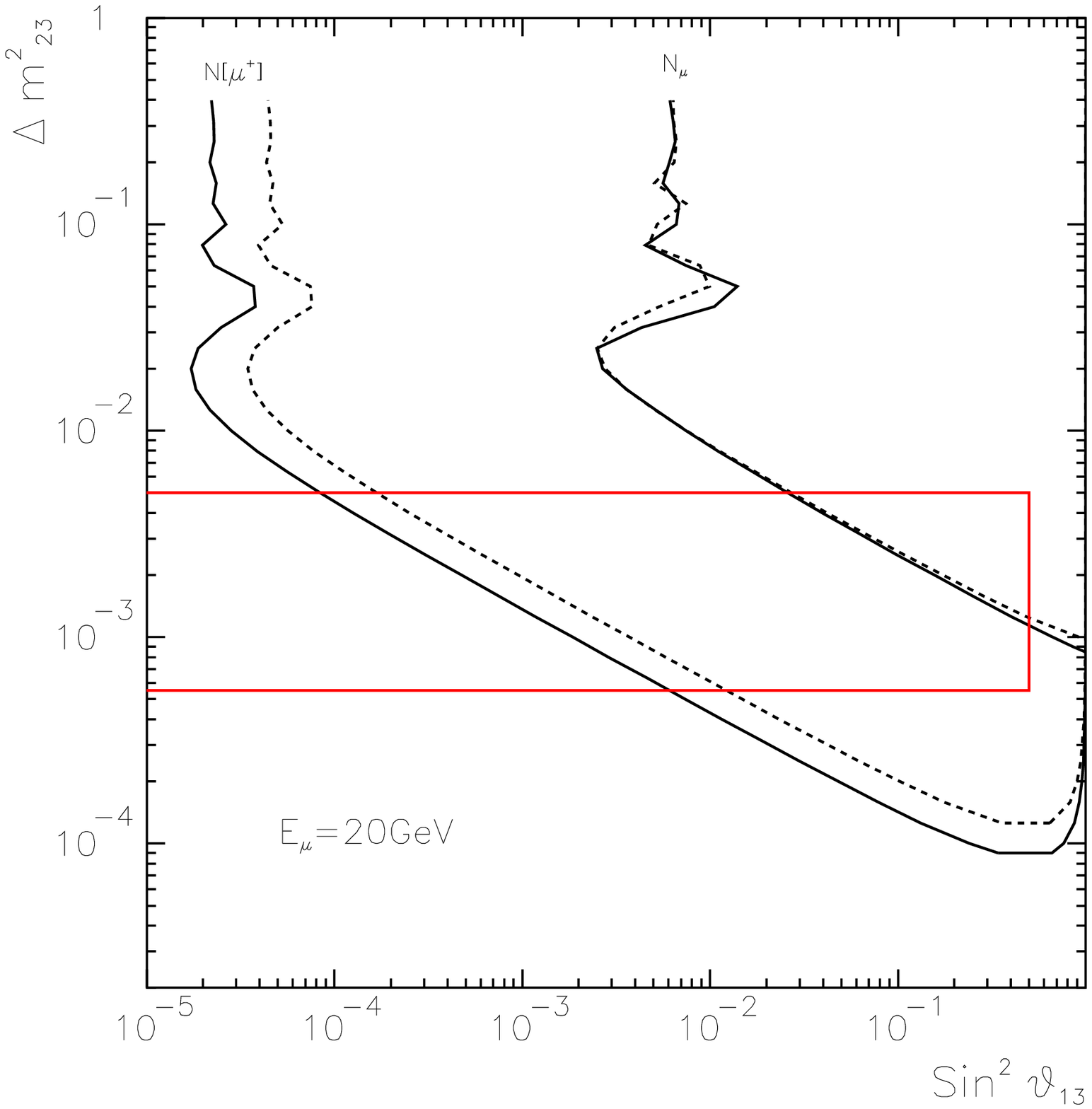,width=4.in,height=4.in}}
\caption[]{Sensitivity reach in the plane 
$[\sin^2 \theta_{13},\Delta m_{23}^2]$, at 90\% confidence, for the same conditions as in Fig.~\ref{fig:t23m}.
 The continuous (dashed) lines correspond to
$\theta_{23}=45^0\, (30^0)$.
 The lines covering the most 
(least) ground are for the appearance (disappearance) observable $N[\mu^+]$ ($N_\mu$). The rectangular domain is the approximate region allowed
by SuperK data.}
\label{fig:t13m}
\end{figure}

\begin{figure}[htb]
  \centering
\mbox{\epsfig{file=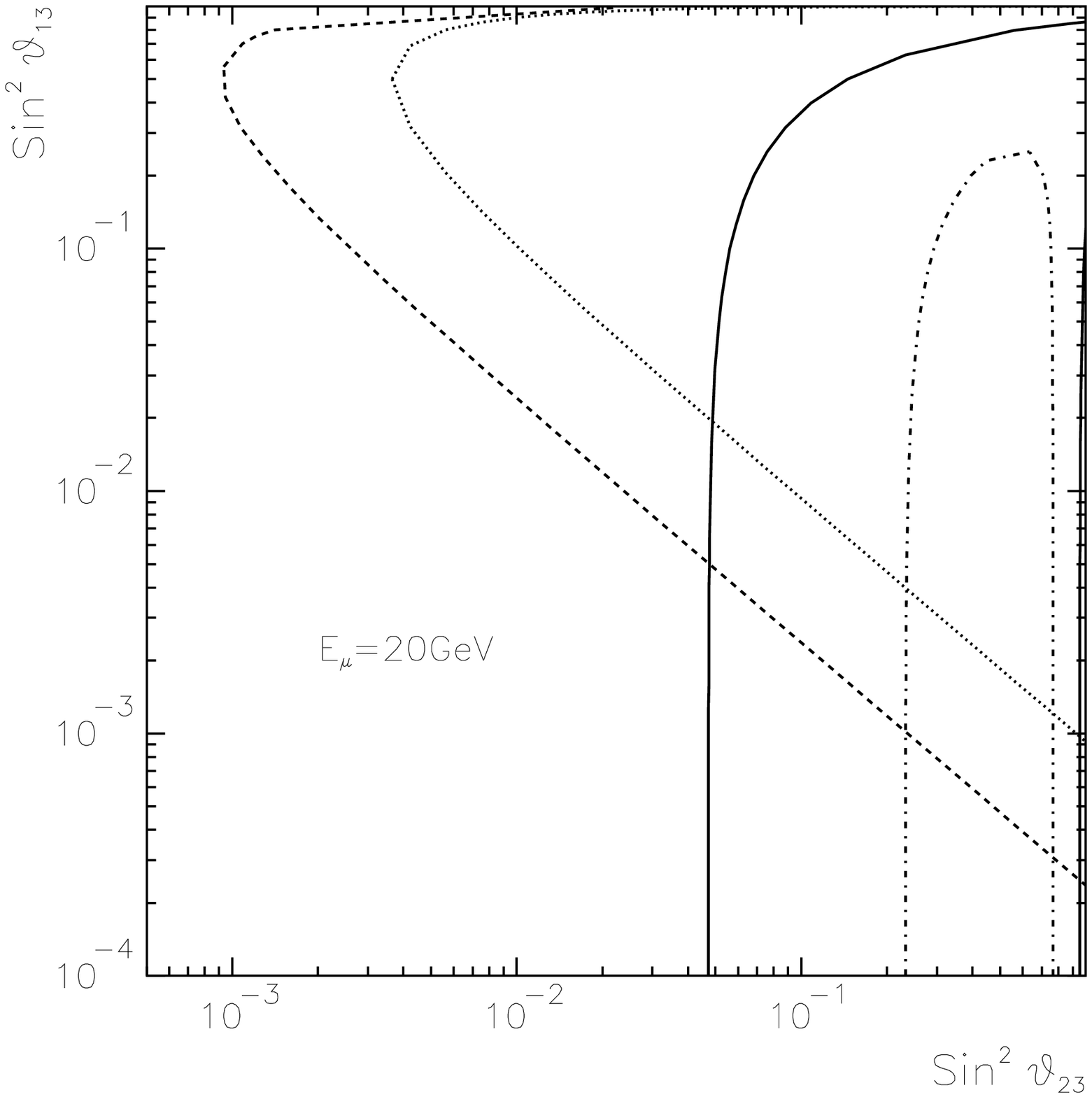,width=4.in,height=4.in}}
  \caption{Sensitivity reach in the
[$\sin^2\theta_{23}, \sin^2\theta_{12}$] plane at 90\% confidence, 
for the same conditions as in Fig.~\ref{fig:t23m}.  The dashed and 
dotted lines correspond to the 
appearance observable $N[\mu^+]$
with  $\Delta m_{23}^2= 2 \times 10^{-3}$ eV$^2$, and  
$\Delta m_{23}^2= 10^{-3}$ eV$^2$, respectively.
The regions interior to the continuous and dot-dashed lines are
exclusion domains stemming from
the disappearance observable, $N_\mu$,
with  $\Delta m_{23}^2= 2\times 10^{-3}$ eV$^2$, and  
$\Delta m_{23}^2= 10^{-3}$ eV$^2$, respectively.
} 
\label{fig:t23t13}
\end{figure}

In Fig.~\ref{fig:t13m} we show the sensitivity reach, in the 
[$\sin^2(\theta_{13}),\Delta m_{23}^2$] plane
for the extremal values of 
$\theta_{23}\sim 30^0,\, 45^0$ allowed by the
SuperK data. In Fig.~\ref{fig:t23t13} we show the sensitivity reach
in the plane $[\sin^2\theta_{23},\sin^2\theta_{13}]$.

The overall conclusion of this analysis in terms of the
mixing of three generations is that the capability of detecting
``wrong-charge'' muons would be extremely useful in giving access
to the study of a large region of the 
($\theta_{13}$, $\theta_{23}$, $\Delta m_{23}^2$) parameter space.

\section{Matter effects and scaling laws}

Of all neutrino species, only $\nu_e$ and $\bar\nu_e$ have  charged-current elastic scattering amplitudes on electrons. 
This, it is well
known, induces  effective ``masses'' $\mu=\pm\, 2\,E_\nu\, A$,
where the signs refer to $\nu_e$ and $\bar\nu_e$ and
$A=\sqrt{2}\, G_F\, n_e$, with $n_e$ the ambient electron
number density \cite{MSW}. Matter effects 
\cite{MSW,CPtoda} are important
if $A$ is comparable to, or bigger than, the quantity $\Delta_{jk}= \Delta m_{jk}^2/(2\,E_\nu)$ of Eq.(\ref{Deltas}) for some mass
difference and neutrino energy. In the minimal scheme
$\Delta m_{12}^2$ is neglected relative to $\Delta m_{23}^2$,
the question is the relative size of $A$ and
$\Delta_{23}\simeq \Delta_{13}$ (we assume $\Delta m_{23}=
m_3^2-m_2^2$ to be positive, otherwise the roles of neutrinos
and antineutrinos are to be inverted in what follows). 


For the Earth's crust,
with density $\rho\sim 2.8$ g/cm$^3$
and roughly equal numbers of protons, neutrons and electrons,
$A\sim 10^{-13}$ eV. The typical neutrino energies we are considering are tens of GeVs.
 For $E_\nu=12$ GeV (the average $\bar\nu_e$ energy in the decay of $E_\mu=20$ GeV muons) 
$A\simeq\Delta_{23}$ for 
$\Delta  m^2_{23}=2.4\times 10^{-3}$ eV$^2$.
This means that $A\gg \Delta_{23}$
for the lower $\Delta m^2$ values in 
Figs.~\ref{fig:t23m},\ref{fig:t13m}
while the opposite is true at the other end of the relevant mass domain.
Thus, the matter effects that we have so far neglected are dominant in 
the most relevant portion 
of the domain of interest: the lower mass scales. Yet, as we
proceed to show, 
matter effects are practically irrelevant 
(except in the analysis of CP-violation effects) in long baseline
experiments with $L<3000$ km. They only begin to have a sizeable 
impact at even larger distances\footnote{This refers to the
approximate assessment of sensitivities, not to the analysis of eventual
results: in the Sun or on Earth, Nature may well have chosen parameter values 
for which matter effects are relevant.}.

Define 
\begin{equation}
B\equiv \sqrt{\left[\Delta_{23}\,\cos(2\theta_{13})-A\right]^2
+\left[\Delta_{23}\,\sin(2\theta_{13})\right]^2}
\label{B}
\end{equation}
and
\begin{equation}
\sin(2\,\theta_M)\equiv{\Delta_{23}\,\sin(2\theta_{13})/ B}\, ,
\label{thetamatter}
\end{equation}
where $\theta_M$ is to be taken in the first (second) quadrant
if $\Delta_{23}\,\cos(2\theta_{13})-A$ is positive (negative).
The transition probability governing the appearance of wrong sign
muons is, in the minimal scheme, in the presence of matter
effects, and in the approximation of constant $n_e$ \cite{Yasuda}:
\begin{equation}
P(\nu_e\rightarrow\nu_\mu)\simeq  s^2_{23}\, 
\sin^2(2\theta_M)\,\sin^2\left({B\, L/ 2}\right)
\label{probmatt1}
\end{equation}
which, for $A=0$, reduces to the corresponding vacuum result:
the first of Eqs.(\ref{todasprobs}).
For  $B\, L/2$ sufficiently small, it is a good approximation
to expand the last sine in Eq.(\ref{probmatt1}) and to use
Eq.(\ref{thetamatter}) to obtain:
\begin{equation}
P(\nu_e\rightarrow\nu_\mu)\sim s^2_{23}\, 
\sin^2(2\theta_{13})\,\left[\Delta_{23}\,L/2\right]^2 \, ,
\label{probmatt11}
\end{equation}
which coincides with the expansion for small 
$\Delta_{23}\,L/2=\Delta m^2_{23}\, L/(4\,E_\nu)$
of the vacuum result in Eqs.(\ref{todasprobs}), even when
matter dominates and $B\simeq A$ (at a distance of $L=732$ km, $A\,L/2\sim 0.2$). 

In practice, 
and after integration over the neutrino flux and cross section,
the above approximations are excellent in that part
of the disappearance sensitivity contours of 
Figs.~\ref{fig:t23m}-\ref{fig:t23t13}
that are roughly ``straight diagonal'' lines of slope $-1$. There, 
$s_{23}\, \sin(2\theta_{13})\,\Delta m^2_{23}$ is
approximately constant. In this region the 
results with and without matter effects are indistinguishable
and (for equal number of events) the sensitivity contours from $\nu_e\to\nu_\mu$
and $\bar\nu_e\to\bar\nu_\mu$ transitions would also coincide.

For sufficiently large $\Delta m^2_{23}$, matter effects are negligible.
In Figs.~\ref{fig:t23m},\ref{fig:t13m} this occurs in the portion of
the limits that are approximately ``straight vertical'' lines, for which the
oscillating factors in Eqs.(\ref{probmatt1},\ref{probmatt11}) average
to 1/2. All in all, only the wiggly regions in the
sensitivity boundaries distinguish
matter from vacuum, neutrinos from antineutrinos. The differences
are not large (factors of order two). All of the above also applies to the 
disappearance-channel results shown in the same figures.

The preceding discussion was made in the context of the relatively
``short'' long baseline of 732 km and for $E_\mu=20$ GeV. How do
our results scale to other distances and
stored-muon energies? (the scaling laws differ somewhat from similar
ones for neutrinos from $\pi$ and $K$ decay).

We are considering detectors at a sufficiently
long distance (or otherwise sufficiently small in transverse
dimensions) for the neutrino beam that bathes them to be
transversally uniform. For a fixed number of decaying muons 
(independent of $E_\mu$) the forward neutrino flux 
varies
as $E_\mu^2\,L^{-2}$, see Eq.(\ref{flux}). The neutrino cross sections at moderate energy
are roughly linear in the neutrino (or parent-muon) energy. For
$L<3000$ km, $\sin^2(A\, L/2)\sim(A\, L/2)^2$ is a good approximation 
(better than 25\% and rapidly deteriorating for increasing $L$) 
and the vacuum-like result of Eq.(\ref{probmatt11}) is applicable. 
Entirely analogous considerations apply to the probability
$P(\nu_\mu\to\nu_\mu)$ whose explicit form in the minimum scheme
\cite{Yasuda} we have not written. All this implies that the
``straight diagonal'' parts of the
appearance contours in Figs.~\ref{fig:t23m}-\ref{fig:t23t13}
scale as $s_{23}\, \sin(2\theta_{13})\,\Delta m^2_{23}\propto
E_\mu^{-1/2}$, with no $L$ dependence. For $L>3000$ km, this
sensitivity (still in the approximation of constant $n_e$)
is weakened by an extra $L$-dependent factor so that, for any distance,
the appearance sensitivity at the low-mass end 
scales as:
\begin{equation}
s_{23}\, \sin(2\theta_{13})\,\Delta m^2_{23}\propto
E_\mu^{-1/2}\;(A\, L/2)/\big|\sin(A\, L/2)\big|\; .
\label{sensitivity}
\end{equation}
For the ``straight vertical'' parts of the appearance boundaries
in Figs.~\ref{fig:t23m},\ref{fig:t13m} the oscillation probabilities 
average to 50\% and the scaling law is  $s_{23}\, \sin(2\theta_{13})
\propto L\, E_\mu^{-3/2}$. 

For a disappearance channel the putative signal must compete with
the statistical uncertainty in the background and the $E_\mu$ and $L$
dependence are not those of an appearance channel. Moreover, the
scaling laws for our $N[\mu^+ +\mu^-]$ contours
are not very simple functions of the mixing angles. 
For $L<3000$ km
their ``straight diagonal'' portions
in Figs.~\ref{fig:t23m},\ref{fig:t13m} scale up and down as
$\Delta m^2\propto E_\mu^{1/4}\,L^{-1/2}$.        
The ``straight vertical'' parts of these limits move right and left as
$\sin \theta\propto L^{1/2}\, E_\mu^{-3/4}$. For $L>3000$ km
the scaling laws for disappearance are more involved.

\begin{figure}[htb]
\centering
\mbox{\epsfig{file=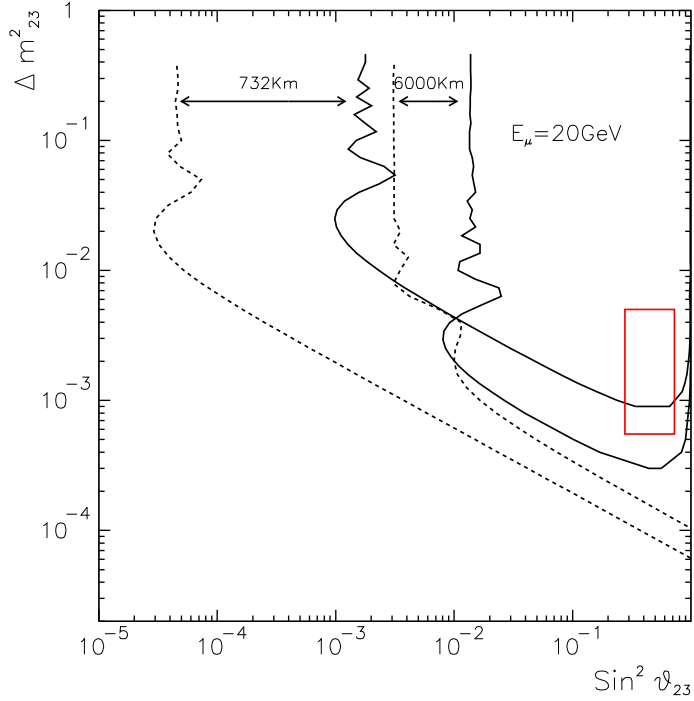,width=4.in,height=4.in}}
\caption[]{Sensitivity reach in the plane 
$[\sin^2 \theta_{23},\Delta m_{23}^2]$
 at 90\% confidence, for our reference set-up, a $\mu^-$-decay
beam and $L=732,\, 6000$ km.
The discontinuous (continuous) lines
correspond to the appearance (disappearance) observable
$N[\mu^+]$ ($N[\mu^++\mu^-]$).
We chose $\theta_{13}=40^0$ for appearance, 
 $\theta_{13}=0$ for disappearance.  }
\label{fig:lejos}
\end{figure}

In Fig.~\ref{fig:lejos} we compare results
for $L=732$ and 6000 km. Only the disappearance
channel at large $\sin^2\theta$ benefits from the larger
distance. For the more attractive wrong-sign $\mu$-appearance 
channel there is no advantage to a very long baseline.

\section{T and CP violation ?}

The beams from a hypothetical neutrino factory would be so
intense and well understood that one may daydream about
measuring CP violation in the very clean environment of a
$\mu$-decay beam. Standard-model CP-violation effects, as is well
known in the quark sector, entail an unavoidable reference to
all three families. They would consequently vanish in the
minimal scheme that we have been considering, insofar as
the mass difference $\Delta m^2_{12}$ is neglected. With the
inclusion of this difference the parameter space (two mass gaps,
three angles, one CP-odd phase) becomes so large that
its conscientious exploration 
would, in our current nescient state,
be premature. We will simply 
give some examples  of the size of the effects that one could,
 rather optimistically, expect.

CP-related observables often involve the comparison between measurements 
in the two charge-conjugate modes of the factory.
One example is the asymmetry \cite{Nicola} 
\begin{equation}
 A_{e \mu}^{CP}\equiv\frac{P(\nu_e\rightarrow \nu_\mu)-
P(\bar\nu_e\rightarrow \bar\nu_\mu)}{P(\nu_e\rightarrow
 \nu_\mu)+
P(\bar\nu_e\rightarrow \bar\nu_\mu)}\,
\label{CPodd}
\end{equation}
which would, in vacuum, be a CP-odd observable.  The voyage through 
our CP-uneven planet, however, induces a non-zero 
$A_{e \mu}^{CP}$ even if CP is conserved, since $\nu_e$
and $\bar\nu_e$ are differently affected by the ambient 
electrons \cite{Arafcp}.

In a neutrino factory $A_{e \mu}^{CP}$ 
would be measured by first
extracting $P(\nu_\mu\rightarrow \nu_e)$ from the produced 
(wrong-sign) $\mu^-$s in a beam from $\mu^+$ decay and 
 $P(\bar\nu_e\rightarrow \bar\nu_\mu)$ from the charge conjugate
beam and process. Even if the fluxes are very well
known, this requires a good knowledge of the cross section
ratio $\sigma(\bar\nu_\mu\to\mu^+)/\sigma(\nu_\mu\to\mu^-)$, which
may be gathered in a short-baseline experiment. To obtain the 
genuinely CP-odd quantity of interest, the matter effects
must be subtracted with sufficient precision. But we shall see that
the truly serious limitation
is the small statistics inherent to appearance channels. 

 The T-odd asymmetry \cite{Rusos}
\begin{equation}
 A_{e \mu}^{T}\equiv\frac{P(\nu_e\rightarrow \nu_\mu)-
P(\nu_\mu\rightarrow \nu_e)}{P(\nu_e\rightarrow \nu_\mu)+
P(\nu_\mu\rightarrow \nu_e)}\; 
\label{CPt}
\end{equation}
is ``cleaner'' than the CP-odd one, in that a non-zero
value for it cannot be induced by matter effects.
As a consequence of  CPT-invariance the two asymmetries,
in vacuum,
are identical
$A_{e \mu}^{T}[{\rm vac}]=A_{e \mu}^{CP}[{\rm vac}]$.
The $T$-odd asymmetry 
is very difficult to measure in practice.
In a $\mu^-$-generated beam
the extraction of $P(\nu_\mu\rightarrow \nu_e)$ would require
a measurement of electron charge, the $e^++e^-$ number involving
also $P(\bar\nu_e\rightarrow \bar\nu_e)$. It is not easy to measure
the electron charge in a large, high-density experiment.

The complete expressions for $A^{CP}_{e \mu}$ in the presence of
matter  are rather elaborate and we do not reproduce
them here. To illustrate the size of the effects,
in Table 1 we give the values of various asymmetries at
$L=732$ km with a fixed neutrino energy, $E_\nu=7$ GeV
 with maximal CP violation, $\delta=90^0$,
and with various parameter values chosen in their currently
allowed domains.
The Table reports the vacuum asymmetry
$A^{CP}_{e\mu}[{\rm vac}]$,  
the calculated expectation $A^{CP}_{e\mu}(0)$
for the apparent CP-odd asymmetry induced
by matter, and the genuine CP-odd asymmetry in matter: 
\begin{equation}
{{\cal A}_{e\mu}}(\delta)=
A^{CP}_{e\mu}(\delta)-A^{CP}_{e\mu}(0) \; ,
\label{CPfixed}
\end{equation} 
in which the matter effect is subtracted.

\vskip .5cm
\centerline{
\vbox{\tabskip=0pt \offinterlineskip
\def\tablerule{\noalign{\hrule}}
\halign to367pt{\strut#& \vrule#\tabskip=1em plus2em
&\hfil#& \vrule#
&\hfil#& \vrule#
&\hfil#& \vrule# 
& \hfil#\hfil& \vrule#
&\hfil#& \vrule#
&\hfil#& \vrule#   
&\hfil#& \vrule#
\tabskip=0pt\cr\tablerule
&&\omit
$\sin^2\theta_{12}$
&&
\omit $\theta_{13}$ 
&&
\omit \hfil $\Delta m^2_{12}$ \hfil 
&&
\omit \hfil $A^{CP}_{e\mu}[{\rm vac}]$ \hfil
&&
\omit \hfil ${\cal A}_{e\mu}$ \hfil
&&
\omit \hfil $A^{CP}_{e\mu}(0)$ \hfil 
&\cr\tablerule
&&   0.5 
&& 13$^0$ 
&& $10^{-5}$  
&&\hfil $-5.9\, 10^{-3}$ \hfil 
&&\hfil $-5.5\, 10^{-3}$ \hfil
&&\hfil $ 1.6\, 10^{-2}$ \hfil
&\cr\tablerule
&&   $5\, 10^{-3}$ 
&& 30$^0$ 
&& $10^{-4}$ 
&&\hfil $-3.4\, 10^{-3}$ \hfil
&&\hfil $-3.2\,10^{-3}$ \hfil
&&\hfil 9.8 $10^{-3}$ \hfil
&\cr\tablerule
&&    0.5  
&& 30$^0$ 
&& $10^{-4}$ 
&& $-2.6\, 10^{-2}$ 
&& \hfil $-2.5\, 10^{-2}$ \hfil
&& \hfil $7.8\, 10^{-3}$ \hfil
&\cr\tablerule
&& 0.5 
&& 13$^0$ 
&& $10^{-4}$
&&\hfil $-5.6\, 10^{-2}$ \hfil
&&\hfil $-5.4\, 10^{-2}$ \hfil
&&\hfil $1.4\, 10^{-2}$ \hfil
&\cr\tablerule}}
}
\vskip.2cm
\noindent{Table 1:  The CP asymmetries defined in the text,
at $L=732$ km, for $\delta=\pi/2$,
  $\theta_{23}=45^0$,
$\Delta m^2_{23}=10^{-3}$ eV$^2$, $E_\nu=7$ GeV and
choices of other parameters compatible
with solar and atmospheric data. 
\vskip.2cm

With no further ado, Table 1 conveys the message that,
if $\Delta m^2_{21}$ is indeed as small as the ensemble of
solar neutrino experiments would imply, the CP-odd
effects are only sizeable in a small domain of  parameter
space, exemplified here by the last two rows of the table. 
Is that region amenable to empiric scrutiny? 

A first question concerns the
relative size of the measured and the theoretically subtracted 
terms. For the subtraction procedure to be useful $\theta_{23}$, $\theta_{13}$, 
$\Delta m^2_{23}$ and the density profile traversed by the beam
must be known with sufficient precision
for the error in the subtracted term not to dominate the result.
At the distance of $L=732$ km used to construct Table 1, this
does not seem to be a problem: for the parameter values
of the last two rows, the subtractions are small enough
that a precision of a factor of two in their determination would
suffice. 

A second question on the observability of CP-violation is that of statistics.
In practice, for our reference set-up, there would be too few events to
exploit the explicit $E_\nu$ dependence of the CP-odd effect. 
To construct a realistic CP-odd observable, consider the neutrino-energy 
integrated quantity: 
\begin{equation}
{\bar A}^{CP}_{e\mu} = \frac{\large\{{N[\mu^-]}/{N_o[e^-]}\large\}_{+} 
- \large\{N[\mu^+]/N_o[e^+]\large\}_{-}} {\large\{N[\mu^-]/N_o[e^-]\large\}_{+}
 + \large\{N[\mu^+]/N_o[e^+]\large\}_{-}}\; ,
\label{intasy}
\end{equation}
where the sign of the decaying muons 
is indicated by a subindex,
$N[\mu^+]$ $(N[\mu^-])$ are the measured number of wrong-sign muons, and 
$N_o[e^+]$ $(N_o[e^-])$ are the expected number of $\bar{\nu}_e (\nu_e)$ 
charged current interactions in the absence of 
oscillations\footnote{ In the analogue
energy-integrated T-odd asymmetry, the T-even contributions to its 
numerator do not cancel,  
due to the different energy distributions of $\nu_e$s and $\nu_\mu$s in 
the beam.}.
The genuine CP-odd asymmetry is
${\overline {\cal A}}_{e\mu}(\delta)=
\bar A_{e\mu}^{CP}(\delta) -\bar A_{e\mu}^{CP}(0)$, the
flux and cross-section weighed version of Eq.(\ref{CPfixed}).

In Fig.~\ref{fig:CP} we give the signal over statistical noise
ratio for $\big|{\overline {\cal A}}_{e\mu}(\pm \pi/2)\big|$ 
as a function of distance for our 
standard set-up, for $E_\mu=10,20$ GeV  
and for the parameters in the last row of Table 1.
The number of  ``standard deviations'' is seen not to
exceed $\sim 2$ at any distance. Moreover, for very
long baselines,  the relative size of the theoretically subtracted
term $\bar A_{e\mu}^{CP}(0)$ increases very rapidly,
as shown in Fig.~\ref{fig:subtract}. 
We have examined other parameter values within the limits
of the scenario we have adopted for neutrino masses and mixing
angles\footnote{The CP-violation effects are much bigger
for the larger mass differences that become possible if
the results of some solar neutrino experiment are disregarded.
 We have not pursued this option.}. As an example,
increasing $\Delta m^2_{23}$ from $10^{-3}$ to $6 \times 10^{-3}$
eV$^2$, with the other parameters fixed as in Table 1, increases
the maximum number of standard deviations to $\sim 3.5$ 
(at $L\sim 3000$ km) but the relative size of the theoretically
subtracted term at that distance increases by an order of magnitude
relative to what it is in Fig.~\ref{fig:subtract}.

The conclusion is that,
if the neutrino mass differences are those indicated by
solar and atmospheric observations and the physics is 
that of three standard families, there is little hope
to observe CP-violation with the beams and detectors
we have described.

\begin{figure}[htb]
  \centering
\mbox{\epsfig{file=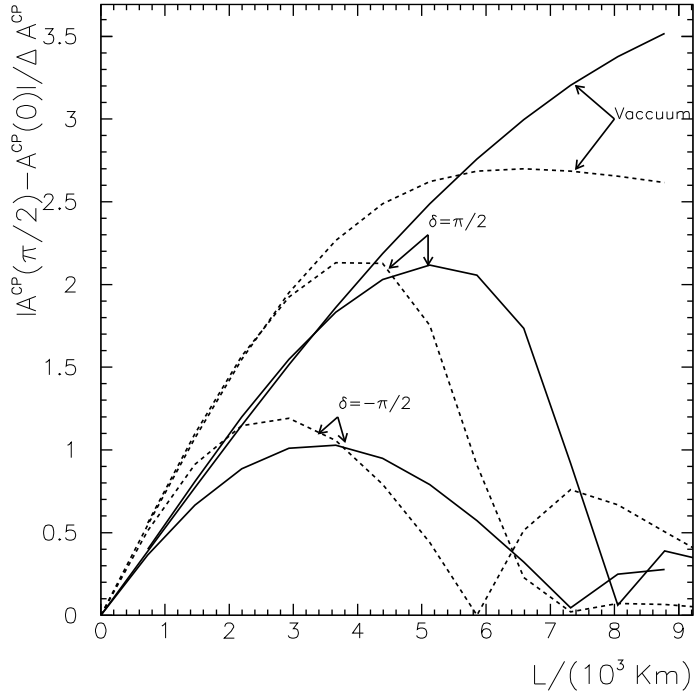,width=4.in,height=4.in}}
  \caption{Signal over statistical uncertainty in a measurement
of CP asymmetries as a function of distance, with the continuous
(dashed) lines corresponding to
$E_\mu=20\, (10)$ GeV. The chosen CKM parameters are
those of the last row of Table 1. The lower four curves describe
$\big|{\overline {\cal A}}_{e\mu}(\pm \pi/2)\big|$ over its statistical
error. The upper two curves are vacuum results for the same
CP phase(s). 
} 
\label{fig:CP}
\end{figure}

\begin{figure}[htb]
  \centering
\mbox{\epsfig{file=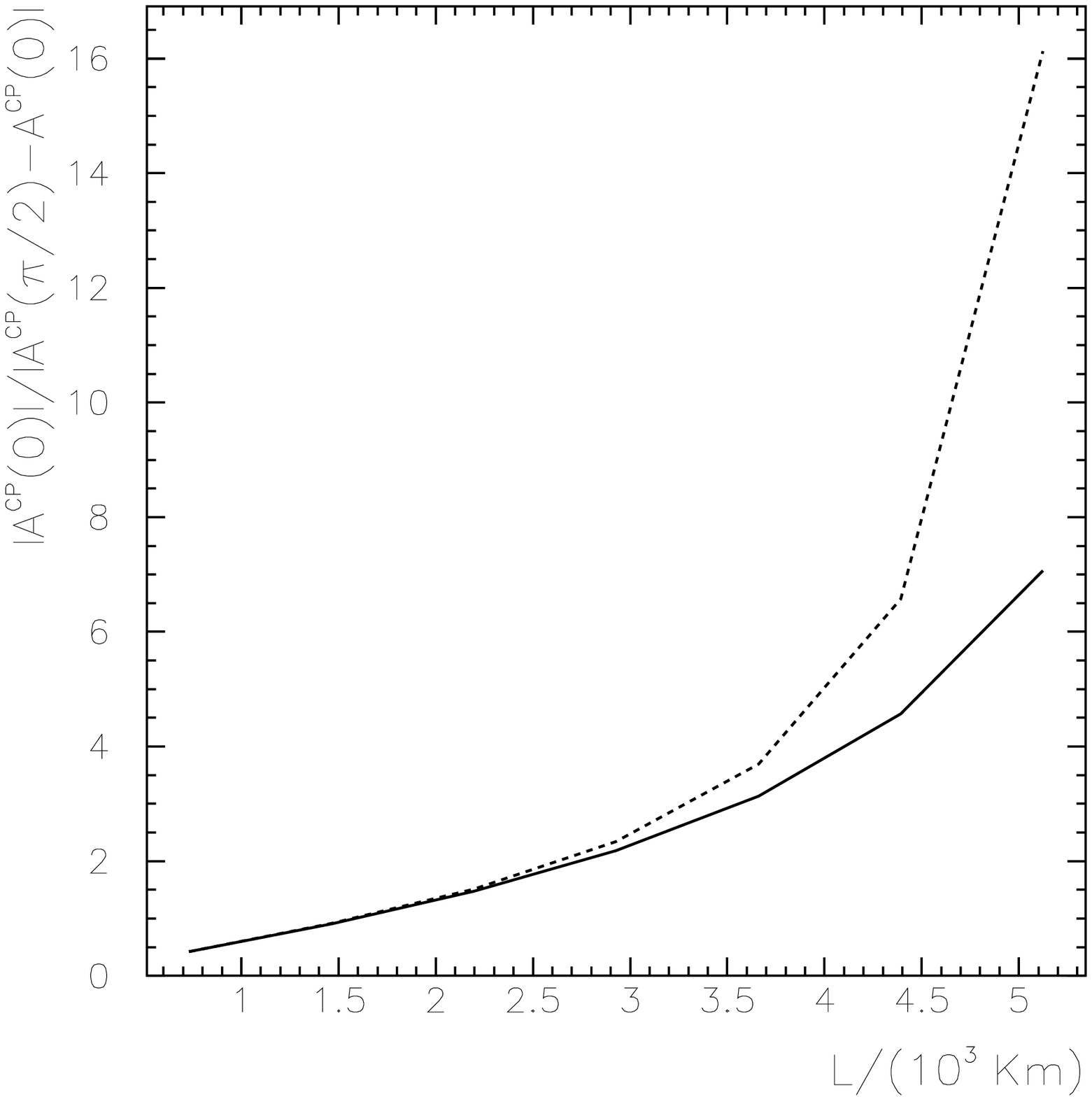,width=4.in,height=4.in}}
  \caption{Ratio of the subtracted term $\bar A_{e\mu}^{CP}(0)$ 
relative to the genuine CP asymmetry ${\overline {\cal A}}_{e\mu}(\pi /2)$,
as a function of distance, with the continuous
(dashed) lines corresponding to
$E_\mu=20\, (10)$ GeV. The chosen CKM parameters, are
those of the last row of Table 1.} 
\label{fig:subtract}
\end{figure}

\section{Observables and
backgrounds in $\pi$- and $\mu$-decay beams.}

In a search for $\tau$ appearance, a $\mu$-decay beam, but for
its conceivable intensity,
would not have overwhelming advantages relative to a conventional
$\pi+K$ decay beam; the contamination of $\nu_\tau$ from $D_s$
decay in the $\pi$ beam is known to be small, witness
the fact that the third generation
neutrino has not yet been ``seen''. The background from the charmed
particles produced by the other neutrino types would be equally
challenging in a conventional or a $\nu$-factory beam. We briefly
compare
these beams for oscillation studies other than $\tau$ appearance. 

The $\nu_\mu$ beams from $\pi$ decay have a contamination
of $\nu_e$s from $K_{e3}$ decays. A small contamination
of the wrong helicity neutrinos (e.g. $\bar\nu_\mu$ in
a predominantly $\nu_\mu$ beam) is also unavoidable, due
to limitations of the charge-separation and focusing
system. It is difficult to understand
these beams theoretically to better than 10\% precision.
With a $\pi^+$ decay beam one can measure neutral currents
and the production of electrons and muons, the measurement
of whose charge is immaterial; that is, a total of three
observables, one of which (electron events) is beset
by background problems. Ideally
beams of opposite polarity add information,
but the comparison of $\nu_\mu$ and $\bar\nu_\mu$ disappearance channels 
for a study of CP-violation would be even more 
demanding than for the $\nu$-factory wrong-sign $\mu$-appearance
examples discussed in the previous section.

The number of useful observables in an experiment with a
 $\nu_\mu+\bar\nu_e$ beam from $\mu^-$ decay is 
larger than for a $\pi$ decay beam.
Assume that one or various aligned experiments are capable of
distinguishing $\mu^+$, $\mu^-$, $e^++e^-$, and neutral current
events. One of these observables ($\mu^+$ appearance) is a tell-tale
signal of oscillations. From the other three observables one
can extract information on oscillation probabilities with
errors associated only with statistics, backgrounds, efficiencies and 
cross sections, but with very small flux uncertainties.  In total,
 for each polarity, a $\mu$-decay facility could measure four channels 
other than $\tau$ appearance. In principle this is sufficient to
determine (or severely constrain) two of
the three Euler angles ($\theta_{23}$ and $\theta_{13}$) 
of the neutrino-mixing matrix in Eq.(\ref{CKM})
and (with a measurement of $E_\nu$) the 
neutrino mass splitting $\Delta m^2_{23}$.
With a conventional $\pi$-decay beam such a
program would be out of reach\footnote{Charged pions and 
kaons decay two orders of magnitude faster
than muons. Only if there was time, in a brief pion lifetime, to clean
up a pion beam of its kaon contamination by some electromagnetic
gymnastics, would a ``pion factory'' compete with a $\mu$-decay 
race-track as a candidate neutrino factory.}.

The backgrounds to a wrong-sign $\mu$ signal are not associated 
with the beam, but with the numerous decay processes that can
produce or fake such muons. Pions masquerading as muons can
be ranged out with great efficiency, particularly in competition
with the generally energetic primary muon from the leptonic vertex.
Muonic charged currents are not the most threatening background,
since one would also have to miss the right-sign muon. 
Electronic charged currents may singly produce charmed 
particles, but the decays of the latter lead to muons of the ``right'' sign. 
In any case, the  
background from charm production and subsequent muonic
decay can be easily suppressed or studied by lowering
$E_\mu$ below the canonical 20 GeV we have been using.  At
1/4 the stored muon energy the statistical appearance sensitivity
would be reduced by a factor of 2, while charm production would be
almost completely kinematically forbidden (this is 
an extreme example, in that it might jeopardize muon recognition). 
Neutral current events
in which a hadron decays into a muon early or straight enough
are presumably the main
hazard. Experience with NOMAD --admittedly not a coarse-grained
very large device-- demonstrates that an `isolation' cut in the transverse
momentum of the muon candidate relative to the direction
of the hadronic jet is extremely efficient
\cite{bcr2}. In these events, an additional cut of the missing 
transverse momentum 
(carried mainly by the outgoing neutrino in the neutral-current
leptonic vertex) relative to the muon plus hadrons would also
help. Even detectors as coarse-grained as MINOS \cite{MINOS}
or NICE \cite{NICE}
have jet-direction
reconstruction capabilities and could implement similar cuts. 

Without a specific detector in mind and
considerable simulation toil we cannot answer the question of
how large the above backgrounds would be. A question that we can
answer is how small they would have to be not to interfere
with the signal. For our standard set-up and an unoptimized
$E_\mu=$ 20 GeV, there would be a grand total of a few $10^5$ 
events for $n_\mu=2\times 10^{20}$ $\mu$ decays at $L=732$ km. 
To compete with a limiting appearance signal of a few 
wrong-sign muons may be difficult. 
At some 10 times larger $L$ the
low-mass edge of the sensitivity domain 
would change very little, as shown in 
Eq.(\ref{sensitivity}) and Fig.~\ref{fig:lejos},
while the background would be reduced by two orders of magnitude,
a level at which it would not represent a challenge.
The overall optimization of the signal-to-noise ratio is
a multi-parameter task that we cannot engage in.

\section{Summary.}

The inevitable conclusion of a description of
atmospheric and solar neutrino data 
as two independent two-by-two neutrino-mixing effects is that
the only hope to corroborate the atmospheric results with
artificial beams is based on
long baseline experiments looking for $\tau$ appearance
or $\mu$ depletion. These experiments would have great difficulty
in covering the parameter space favoured by SuperK. 
If the same data are analysed in a three-generation mixing
scenario, the conclusions are very different:
long baseline experiments
searching for $\nu_e\leftrightarrow \nu_\mu$ 
transitions regain interest,
since these oscillations (even if primarily responsible
for the long-distance solar effect) will in general also occur over the shorter 
range implied by the atmospheric data.

We have studied  $\nu_\mu\leftrightarrow \nu_e$ oscillations
in the context of a neutrino factory. Rather than concentrating
on the $\nu_\mu\to \nu_\tau$ process, the observation of which 
is notoriously difficult, we have outlined the possibilities
opened by experiments searching, not only for an
unexpected $e/\mu$ production ratio, but very preferably 
for the appearance of ``wrong
sign'' muons\footnote{In principle, but not in practice,
the search for wrong-sign $e$s would be equally useful.}:
$\mu^\pm$s in a beam from decaying $\mu^\mp$s.
We have not dealt in detail with the problem of backgrounds.

A neutrino factory
may provide beams clean and intense enough, not only to
corroborate the strong indication for neutrino oscillations
gathered by the SuperK 
collaboration, but also to launch a program of precision
neutrino-oscillation physics. 
The number of useful observables is sufficient to determine
or very significantly constrain
the parameters $\theta_{23}$ and $\theta_{13}$ 
and $\Delta m^2_{23}$ of a standard three-generation mixing 
scheme. Only if the neutrino mass differences
are much larger than we have assumed would
a neutrino factory serve to measure the remaining mixing
parameters of the very clean neutrino-mixing sector. 

It is instructive to compare the current programs to measure the CKM
mixing matrices in the quark and lepton sectors. Considerable
effort is being invested, sometimes in duplicate, to improve our
knowledge of the quark sector case, mainly via better studies
of $B$-decay. Even though non-zero neutrino masses are barely
established, the neutrino sector of the theory can be convincingly
argued to herald physics well beyond the standard model \cite{Wil}.
It is in this perspective --with dedicated $B$-physics experiments  
and beauty factories in the background-- that a neutrino 
factory should be discussed.

All by itself,  as part of a muon-collider
complex or even as a step in its R\&D, a neutrino factory
seems to be a must.

\section{Acknowledgements}

  We acknowledge useful conversations with B. Autin, 
L. Camilleri, L. Di Lella,
 J. Ellis, J. G\'omez-Cadenas, O. Mena, P. Picchi, F. Pietropaolo, 
C. Quigg, J. Steinberger, P. Strolin and J. Terr\'on. 
M. B. G. thanks the CERN Theory Division for
 hospitality during the initial stage of this work; her work was partially
 supported as well by CICYT project AEN/97/1678.

\end{document}